\documentclass[12pt,a4pa per]{article}

\begin{document}

\title{Source Field Effects and Wave Function Collapse}
\author{ Atilla Gurel atillag@physics-qa.com \\
Hektas Ticaret TAS GOSB Gebze Kocaeli Turkey
 \and Zeynep Gurel zgurel@marmara.edu.tr\\ Marmara
University,Faculty of Education, Physics department,\\
Fikirtepe Istanbul Turkey } \maketitle

\textbf{ }

\textit{ To E.T. Jaynes, in memoriam. } \textbf{ }

\textbf{ }

\textbf{ }
\begin{abstract}
Detection  of a material particle is accompanied by emission of
bremsstrahlung. Thus the dynamics of the energy loss of the
particle is determined by radiation reaction force. The
description of radiation reaction is a difficult problem still
being subject of ongoing debates. There are problems of runaway
solutions, preacceleration already in classical description of
radiation reaction. Additional complications in quantum mechanical
description arise because of the infinite source field energy
term in hamiltonian for a point charge. There is still no general
consensus on an appropriate quantum mechanical description.
Neither the achievements of the radiation theory on the subject
nor the problems associated with it are sufficiently taken into
account in context with measurement problem. Radiation reaction
doesn't effect free particle wave packets, but it favors
stationary states of the `wave function of the measured particle"
in presence of a potential gradient. We suggest therefore that
radiation reaction may play a significant role in the dynamics of
the wave function collapse.

keywords: wave function collapse,interpretation,randomness,
Jaynes Cummings dynamics, spontaneous emission, source field
effects, radiation reaction, quantum measurement,decoherence

\end{abstract}

 \pagebreak

\section{Introduction}

\subsection{Current situation on quantum measurement problem}

There are basically two  different viewpoints (\textbf{1.1.a }and
\textbf{1.1.b}) regarding the problem of wave function dynamics
during a quantum measurement.

\textbf{1.1.a }\textit{``There is no problem at all. The wave
function is not a physical entity in classical sense but it is
only a mathematical construction that allows us to calculate the
probability to observe a particular outcome in a measurement. It
represents therefore not "actual reality" but only ``potential
realities". The ``collapse" therefore is not a continuous time
evolution of a physical entity but only a ``sudden(?)"
actualization of one of these potential realities.  The dynamical
laws of physics (QM) describe only the time evolution of this
probability amplitude but not its ``collapse". Therefore, there is
no sense in looking for a dynamical description of this process.
We have to accept this new reality however discomforting  it may
be philosophically.
 "}

This mental attitude is expressed most clearly by R.P. Feynman
who said in one of his popular lectures (Dudley J.M et al.
(1996)):

\textit{"You will have to accept it. Because it is the way nature
works. If you want to know the way nature works, we looked at it
carefully. Looking at it, that is the way it looks. You don't
like it? Go somewhere else. To another universe, where the rules
are simpler, philosophically more pleasing, more psychologically
easy. I can't help it,okay ?"}

\textbf{1.1.b }\textit{``There is a problem. The measuring
device  itself is obviously nothing but a physical system. Which
 aspect is so fundamentally different in the interaction between
particle and measuring device so that  laws of dynamics of quantum
mechanics (that preserve the superpositions of wave function) are
interrupted during this interaction so that all components in the
initial superposition should vanish except the actual outcome."}

If an electron beam goes between two plates of a capacitor, it
interacts with the electrostatic field. The beam changes its
direction, but there is no collapse. We can verify this by the
fact that the capability of interference of the beam is still
there after it comes out of the capacitor. Ultimately, what
happens in a measuring device is nothing but the interaction of
the beam with the electromagnetic field of the atoms. Thus,
basically there is no qualitative difference between the two
processes but only a quantitative difference regarding the number
of interacting particles, the strength of the electromagnetic
field etc. Thus, there is no reason that something qualitatively
different like ``actualization of one of the potential
realities"  should occur only in one process but not in the other.
Notice that the problem of Copenhagen interpretation is not the
supposedly fundamental randomness, but the problem is the lack of
physical or objective criteria that allows us to determine under
which physical circumstances "one of the potential realities
becomes actualized". Thus, in our opinion, there is indeed a
problem. With all respect to Feynman's creativity and his
extraordinary mind that made great achievements in physics
possible, his words cited above are unacceptable because they
ignore the inherent potential in science for continuous progress.
There is no philosophical argument that justifies the idea ``what
is not known now, should remain unknown for ever".

The suggested solutions to the problem can be classified into two
groups(\textbf{1.1.c }and \textbf{1.1.d}).

\textbf{1.1.c } Some physicists try to solve the problem by
importing new elements or mechanisms into the theory. Bohmian
trajectories(Bohm,(1993)), spontaneous collapse hypothesis of
Girhardi Rimini and Weber (1986), and all hidden variable theories
in general belong to this group.

\textbf{1.1.d } Other physicists try to find solutions within
known QM without adding new elements or mechanisms to the known
equations. Decoherence theories of measurement(Zurek(1982)),
relative state interpretation(Everett,(1957)), and consistent
histories approach (Griffiths (1984))  belong to this group. They
all seem to converge into a single picture of decoherent histories
approach that J.Bub calls "the new orthodoxy"(Bub (1997)page 207).

In this paper we will not discuss the alternative theories
mentioned under \textbf{1.1.c}, but our discussions will remain
within the boundaries of the established theory.

Decoherence\footnote{the vanishing of the off diagonal elements
of the density matrix in pointer variable basis} seems to explain
why we do not observe a superposition of the macroscopically
distinct states of the measuring device corresponding to
different values of the pointer variable in an individual
measurement\footnote{the term decoherence is a very general term
referring to loss of interference. It is sometimes used also to
refer to the destruction of the interference pattern that emerges
as a result of large number of measurements. For example, the
interference pattern on the scintillating screen is lost when
electrons are disturbed while passing through the slits. This
should not be confused with the decoherence between the collective
states of the measuring device  representing the pointer variable
in an \textit{individual} measurement}(Zurek(1982)). However, it
does not explain at first sight why the observed outcome is the
actual one and not another one. This problem is
\textit{``solved"} by interpreting each diagonal element as
parallel evolving but noninteracting branches of classical
reality so that in each branch it appears as if only one
particular outcome is actualized. What we experience as actual
``we" is only one of these branches. Thus the decoherence is
considered as the intrinsic mechanism that leads to branching of
histories that was suggested by Everett.  There is still no
consensus whether the decoherence mechanism is sufficient to
explain the  unique outcome in a measurement or not  (Bub
(1997)page 231, Adler (2003) ). A detailed discussion of the
arguments on this subject is beyond the scope of this paper;
however we want to present one argument as to why we think that
decoherence does not explain the unique outcome in a measurement:

The more the wave function of a particle is localized in space,
the broader its Fourier spectrum is;namely the less localized it
becomes in momentum space. We experience this experimentally as
the uncertainty principle  between position and corresponding
momentum. According to the decoherence picture, on the other hand,
the measurement of position doesn't mean an actual localization of
particle wave function in position space, but it is merely a
slicing of measuring device states each corresponding to detection
at different position into noninterfering branches of classical
reality. Thus, even after the measurement, the state in the total
configurations space (that contains the measured particle and all
particles that make up the measuring device) continues to contain
all positions of the measured particle that were held by the
particle immediately prior to measurement. If this were correct,
measurement of transversal position with a detector would have no
effect on the sharpness of the transversal momentum of the
particle. This contradicts the experience. A measurement of
transversal momentum immediately after position measurement on a
large number of identically prepared particles \footnote{prepared
with sufficiently sharp momentum prior to measurement (plane
wave)} would indeed show that the transversal momentum of the
particle would fluctuate randomly in accordance with Heisenberg
uncertainty principle when it comes out of a thin detector after
being detected. This shows that detection is not merely a slicing
of the total wave function, but that it is an actual localization
of the particle wave function. One can find arguments to overcome
this difficulty but it takes us further and further away from
beautiful simplicity of the known foundations of QM.

One may ask whether known quantum mechanics provides a mechanism
that actually destroys the initial superposition? The general
argument used to rule out this possibility is:

\textit{``This is principally not possible because QM is linear
and the conservation of superpositions is the direct consequence
of this linearity"}

Quantum measurement is a QED problem. In QED the Hamiltonian
contains not only the energy of the particle, but also the energy
of the electromagnetic field; and the wave function in QED is not
the ``particle wave function" of the ordinary QM anymore but a
more abstract wave function that describes the whole system
consisting of material particle(s) and electromagnetic field. In
fact, the term ``particle wave function" is not appropriate
because this wave function cannot be written as a product of two
wave functions because of the correlated states between
electromagnetic field and particle. We will continue to use the
term ``particle wave function" in quotes for practical purposes.
We will discuss in subsection 3.2 and in paragraph 4.a in section
4 why this linearity argument is incorrect  if the subject of
interest is not the time evolution of the whole system but the
time evolution of a subsystem like ``wave function of measured
particle" (as it is the case in a measurement)or more correctly
expressed the time evolution of particle position operator in
Heisenberg picture if we consider the source field of the
particle in total hamiltonian.

\subsection{Radiation reaction and quantum measurement}
In quantum mechanical terms, the detection of an electron is an
inelastic scattering process in presence of large number of
scattering centers accompanied by emission of bremsstrahlung  and
the transversal localization of ``particle wave function". The
scattering theory focuses on differential cross section of
emitted bremsstrahlung based on incoming and outgoing electronic
momentum eigenstates. (Itzykson, C.;Zuber,JB. p.238 (1985)) It
neither focuses on the very mechanism of the energy dissipation
as a consequence of source field dynamics during the short time
period while interacting with the scattering center, nor does it
focus on the effect of this process on the spatial transversal
form of the wave function.

The dynamic localization of the particle wave function without
explicit presence of radiation field is not something completely
unknown to quantum mechanics. Consider, for example, the
transition of an atom from an excited state in discrete spectrum
region to the ground state by spontaneous emission. The spatial
extension of the initial electronic state can be arbitrarily
large if its energy quantum number n is very large.  The role of
radiation reaction in QED description of spontaneous emission was
first realized in the seventies (Ackerhalt et.al.(1973)) and an
intensive discussion regarding the respective roles of vacuum
fluctuations and radiation reaction followed. (Milonni et.al
(1973); Senitzky (1973); Milonni Smith (1975); Milonni (1976)).
Milonni (1976) (1980). According to the currently accepted
picture, spontaneous emission is triggered by vacuum fluctuations
and proceeds under combined effect of vacuum field and radiation
reaction (Dalibard et. al. (1982)(1984); Cohen-Tannoudji (1986)).
Considering the zero expectation value of transversal momentum
and the obvious role of radiation reaction in the detection
process, one wonders whether the transversal localization during
detection may have some aspects common with spontaneous emission
so that there may be no need to import an additional ad hoc
spontaneous localization mechanism as suggested by Girhardi
Rimini and Weber(1986). A problem is that there is still no
general consensus on an accurate quantum mechanical description of
radiation reaction so that it can be applied to a realistic model
of quantum measurement. Discussions and controversies on the
subject continue. Rohrlich (2001)(2002) Ribari\u{c} (2002) Baylis
(2002) O'Connel (2003) Heras(2003)). It is a complicated subject
already in classical description because of problems of runaway
solutions and preacceleration. The problem becomes more
complicated in quantum mechanics if one tries to solve it starting
from first principles because of infinite source field energy in
hamiltonian for a point charge. The price we have to pay to
overcome this difficulty by renormalization in perturbative QED
is that the effect of self retardation is lost in description
(Grotch et.al.(1982)).

Considering the mentioned difficulties related to quantum
mechanical description of source field effects, it is clear that
realistic modeling of a quantum measurement is a very difficult
problem to solve. Current decoherence models of measurement are,
on one hand, far from reflecting this complex aspect of the
process because they ignore source field effects, but on the other
hand, they complicate the problem in a wrong direction by
including the macroscopic measuring device with its huge number
of particles into the total wave function and going to a
configurations space with huge dimensions. This is somewhat a
misleading model, for example, for position measurement with a
scintillating screen, where the real problem gets out of focus.
The significant role of the screen is merely providing a strong
electrostatic field of a particular (quasi-periodic)form that
decelerates the particle so that it emits Bremsstrahlung. It is
the emission of this radiation from a localized spot that makes
up the measurement, and it is the high initial energy of the
particle that leads to the high intensity of the emitted
radiation and to the macroscopicality, thus to the observability
of the effect and not the motion of the pointer of a fictious
macroscopic measuring device.

The following  points are certain:

\textbf{1.2.a }Energy exchange with radiation field  without
explicit presence of external radiation field can proceed  only
under the influence of two factors in QED: Vacuum fluctuations
and/or source field effects.

\textbf{1.2.b }Detection is essentially an energy exchange
between material system and radiation field without explicit
presence of external radiation field(except thermal background
fluctuations and vacuum fluctuations). Detection has not occurred
unless energy exchange with radiation field has occurred.

The consequence of \textbf{1.2.a }and \textbf{1.2.b }is : If it
should be possible at all to understand collapse\footnote{we mean
not a ``slicing" as suggested by decoherence mechanism but an
actual collapse}within laws of QED, without introducing new
elements or mechanisms, it must be a consequence of fluctuations
and/or source field effects.

\subsection{The organization of this paper}

1.In section 2 we introduce the problem of radiation reaction in
quantum mechanics, and we try to outline why do we think that the
consideration of radiation reaction on the dynamics may have a
potential value to open new perspectives for understanding the
mechanism of collapse.

2.In section 3 we go to the history of quantum mechanical
description of radiation reaction to understand why radiation
reaction has not been discussed sufficiently in context with
quantum measurement despite its relevance. In section 4 we
summarize possible reasons.

3. Any successful dynamical collapse theory (based on whatever
mechanism) would make the probability amplitude interpretation
superfluous and would automatically impose a realistic
interpretation of wave function. Therefore, proposing any type of
dynamical collapse mechanism requires  the discussion of the
arguments that are raised against any type of realistic
interpretation and against any type of dynamic collapse approach
in general. This is done in section 5.

\section{Radiation reaction in quantum mechanics}
\subsection{The problem} We will introduce the problem  of
radiation reaction from the perspective of ordinary QM without
intentionally first going to a QED hamiltonian. The reasons will
be clear later.

Schroedinger equation for a particle with electric charge $e$ and
mass $m$ in an external electrostatic potential V(x,y,z) has the
form :
\begin{equation}
-\frac{\hbar^{2}}{2m}\nabla^{2} \Psi(x,y,z) + e \textrm{V}(x,y,z)
\Psi(x,y,z,t) = i\hbar { }\frac{\partial}{\partial t}
\Psi(x,y,z,t)
\end{equation}
using the Ansatz
\begin{equation}
\Psi(x,y,z,t) = \psi(x,y,z) e^{-i\omega t}
\end{equation}
one can obtain the eigenvalue equation
\begin{equation}
-\frac{\hbar^{2}}{2m}\nabla^{2} \psi(x,y,z) + e \textrm{V}(x,y,z)
\psi(x,y,z)= \hbar \omega \psi(x,y,z)
\end{equation}
Equation (3)  has an infinite number of solutions $\psi_{k}$ with
corresponding $\omega_{k}$ such that $E_{k} = \hbar \omega_{k}$ is
the energy of each particular solution. These solutions are
stationary, which means that $|\Psi|^2$  is time independent. The
most general solution of equation (1) can be written as
\begin{equation}
\Psi(x,y,z,t) = \sum_{k} a_{k} \psi_{k}(x,y,z)  e^{-i\omega_{k} t}
\end{equation}
where  $a_{k}$  are constant. The $a_{k}$  satisfy the
normalization condition:\linebreak  $\sum |a_{k}|^2$ = 1 (The sum
should be replaced by an integral for continuous energy regions in
the spectrum). The time evolution  of the expectation value of
the position $ \langle r \rangle   = \int \psi^* \textbf{r} \psi
\textbf{ }dr$ of a localized  wave packet that can be constructed
using (4), is given by Newton's equation of motion
\begin{equation}
m \textrm{ }\frac{d^2}{dt^2} \langle \textbf{r} \rangle = - grad
\textrm{ } V
\end{equation}
This correspondence with classical mechanics is known as the
Ehrenfest theorem. An important property of the solution (4) is
that, since $a_{k}$ are time independent in a static potential,
the expectation value of particle's energy is conserved, namely:
\begin{equation}
\frac{d}{dt} \langle E \rangle_{particle} =\frac{d}{dt}
 \sum_{k} |a_{k}|^2
E_{k} = 0
\end{equation}

There is no dissipation of energy in this model. Classically,
however, accelerated charges radiate and the energy of the
particle decreases because of radiation. One may expect that this
should also occur for an accelerated  quantum mechanical wave
packet. That this expectation is justified is verified by
synchrotron radiation. Thus, obviously eq. (1) is sufficient only
as long as the energy dissipation rate by radiation reaction is
negligible .

\subsection{Potential relevance for quantum measurement}
Although it is clear that radiation reaction is involved in
measurement, we may ask: ``Are there any indications that its
consideration may have any potential value for a better
understanding of collapse or can it provide at most only a more
refined description of the process but without providing any
clues for how collapse occurs"

\textbf{ 2.3.a }Since there must be an energy dissipation from
particle to electromagnetic field in the case of accelerated wave
packet eq(6) cannot be valid. To allow this energy dissipation,
the coefficients $a_k$ in (4) must be time dependent; thus the
solution cannot be written in form of (4) with constant $a_k$.
While the superposition of the form (4) representing a free
particle wave packet (outside the measuring device) is not
effected by radiation reaction, since there is no acceleration ($d
\langle r \rangle /dt = const.$), it begins to decay to lower
energies if it enters a region with a nonzero potential
gradient(a measuring device) because of energy dissipation by
radiation even in the absence of external radiation field! Thus it
is obvious that the initial weights of the particle energy
eigenfunctions in the superposition  are not conserved. Thus
radiation reaction destroys the superposition of ``particle wave
function" that enters the measuring device. This is good news and
shows why a simple ``linearity of QM" argument as mentioned at the
end of the subsection 1.1 is incorrect. This is, however, not
enough because  what we need is not merely the nonconservation of
the weights in the superposition, but what we need is that some
states are preferred so that they can be chosen during
measurement.

\textbf{ 2.3.b }We may suspect that stationary states of eq.(1) in
presence of a potential gradient (like bound atomic energy
eigenstates in detector) should not be affected by radiation
reaction because $d \langle r \rangle /dt = 0$.  Since there is
no truly static field in a real world we may say,  the closer the
particle is to a stationary state of the quasi-static potential
of measuring device the smaller the radiation reaction force is on
the ``particle wave function"  so that these constitute
(relatively) stable islands in the function space of particle.

The consideration of  \textbf{ 2.3.a } and \textbf{ 2.3.b}
together indicate that if the wave function  comes somehow close
to a quasi stationary state during the dissipative time evolution,
it will stay there much longer because of minimal radiation
reaction when compared to a state in full superposition. Thus, the
likelihood to be found in such a quasi-stationary state is
greater then the likelihood to be found in a full superposition.
This indicates that it may be worth studying its possible role in
collapse.

There are of course spatially extended stationary electronic
states in a static potential created by an array of atoms so that
one may doubt whether the above considerations may provide an
understanding of localization. These are however only
theoretically stationary in an ideal case of perfectly static
potential. In a real world the relative positions of atoms
fluctuate because of thermal vibrations so that such spatially
extended electronic states are practically nonstationary and
unstable because of  radiation reaction. It is conceivable
therefore that the electron would become localized in the
energetically most favorable location under these conditions.

To prevent a misunderstanding for a reader with a background in
QED and radiation theory we want to make some remarks in advance.
One may ask ``Is this expectation in accordance with QED or is
this a semiclassical assumption that uses implicitly the already
refuted Schroedinger interpretation of the wave function where $
|\Psi|^2$ = charge density ?". As we will see in subsections 3.4
to 3.7, the presence of radiation reaction in presence  of a
potential gradient doesn't rely on a semiclassical assumption, but
it is the direct consequence of time evolution for particle
position operator in Heisenberg picture if one takes the source
field dynamics into account without leaving the operator
formalism of QED and without using extra semiclassical
assumptions.

Before exploring alternatives like modifying Schroedinger
equation (SE) by nonlinear or ad hoc stochastic terms, it is
important to be aware that SE as it is, neither includes the
effect of vacuum fluctuations that can be a possible candidate
for providing the necessary stochasticity naturally from within
QED, nor does it include the effect of radiation reaction on
particle wave function. Thus vacuum field and source field
dynamics may already provide the necessary modification to SE
naturally. In consideration of the principle of Ockham's razor it
is obviously preferable to look for a solution within QED before
considering ad hoc modifications to SE. These alternatives should
be explored  only if the possibilities within QED are fully
exhausted and proven to be not a solution to the problem.

To understand why the the role of radiation reaction has not been
considered in context with the dynamics of wave function collapse
until now despite these indications, let's take a look at the
history of quantum mechanical description of radiation reaction
that is full of controversies on the interpretation of wave
function.

\section{History: Relation between the quantum description of radiation
reaction and the interpretation problem}

\subsection{Radiation reaction: a problem that is still unsolved classically}

Classically the effect of the energy loss by radiation of an
accelerated charge on the motion of the charge is described by
the well known Abraham-Lorenz equation:

\begin{equation}
m_{eff} \textrm{ }\frac{d^2}{dt^2} \textrm{r} = \textrm{F}(t)
+\frac{2e^2}{3c^3}\frac{d^3}{dt^3} \textrm{r}
\end{equation}

where $m_{eff}$ is the effective mass $m_{eff} = m + \delta m $
and $\delta m $ is the contribution of electromagnetic self
energy to the effective mass. $\delta m $ depends on the size of
the charged particle. The second term on the right describes
radiation reaction and acts similar to a friction force.  There
is, however, a problem with the equation (7). It allows runaway
solutions, namely, solutions where the acceleration increases
exponentially even in the absence of external forces. If one
tries to eliminate the runaway solutions by the imposition of a
suitable boundary condition for $ t\rightarrow \infty$, one
obtains the acausal behavior known as preacceleration.  The
particle accelerates namely before the force acts. The problem
doesn't disappear in relativistic treatment of Dirac either.

\subsection{Things become even more complicated in quantum
description}

We may ask now, for example, how we should modify equation (1) so
that a quantum mechanical equivalent corresponding to the second
term on the right side of eq. (7) automatically emerges on the
right side of the equation (5).

If we want to develop a quantum description, the  problems we face
are:

\textbf{3.2.a} How do we avoid infinite self energy and runaway
problems?

\textbf{3.2.b }We have only time evolution of the particle wave
function, but no definite trajectory. How do we calculate the
changes in the source field if we don't know how the source
particle precisely moves?

There has been basically two different approaches to the source
field problem in the history of quantum mechanics(\textbf{3.2.c}
and \textbf{3.2.d}).

\textbf{3.2.c.} Quantum electrodynamics (QED) includes the energy
of the electromagnetic field in the hamiltonian operator as in
eq.(10). This process is called quantization. Including source
field energy into the hamiltonian in this way requires that it
has to be written as an integral over $E^2$. This presents
difficulties for a pointlike charge because it leads to infinite
self energy. The fact that the runaway solutions are eliminated
for particles with a sufficient size was not known until the
seventies( Moniz and Sharp (1974); Levine et al (1977)). QED
expects that if source field energy is included in hamiltonian,
quantum mechanical description of radiation reaction must follow
automatically.

\textbf{ 3.2.d.} Semiclassical theories (SCT) describe material
particles quantum mechanically, but to avoid the infinite self
energy problem and as a simple solution to problem \textbf{3.2.b}
they use the Schr\"{o}dinger interpretation (SI) of particle wave
function (where $\Psi$ is  not merely probability amplitude but
$|\Psi|^2 =$ classical charge density). Electromagnetic field is
not quantized (E and B  are not operators but c-number fields) and
the internal dynamics of electromagnetic field is described by
Maxwell equations. Thus we have a system of coupled nonlinear
equations that determine the dynamics of the whole system.

\subsection{The first steps in QM: Source field in semiclassical
and neoclassical radiation theory (NCT)}

Because of the difficulty of infinite source field energy for a
pointlike particle, the quantization of the electromagnetic field
was applied only to free field, and the source field problem
couldn't be attacked by QED approach for a long time. In 1927, two
papers appeared both describing the radiation of an atom but
representing the mentioned two different viewpoints(3.2.c and
3.2.d). Dirac took Fourier spectrum of the electromagnetic field
and used the time dependent perturbation theory to calculate its
effect on the atom. Fermi used the Schr\"{o}dinger interpretation
of the wave function and added a nonlinear term containing the
expectation value of dipole moment operator describing the energy
dissipation by radiation reaction. Fermi's model is discussed  by
Wodkiewicz (1980). Fermi found that an excited two level atom
would decay to the lower state with a time dependence that has
the following form

\begin{equation}
|a_{2}(t)|^2  = \frac{1 -|a_{1}(0)|^2} {1 - |a_{1}(0)|^2 +
|a_{1}(0)|^2e^{At} }
\end{equation}

where $a_{2}(t)$ is the weight of the excited state  $\psi_{2}$
and $a_{1}(t)$ is the weight of the lower state $\psi_{1}$ in the
superposition. For $a_1(0) = 0$ there is no decay. For $a_1(0) >
0$ the  decay starts with an initial slope that depends on
$a_{1}(0)$ (small for small $a_{1}(0)$), it reaches a turning
point (the smaller $a_{1}(0)$, the longer it takes) and reaches
asymptotically $e^{-At}$ after the turning point. Interestingly,
the decay rate $A$ in Fermi's formula (8) that determines the
exponential tail (long term behaviour) agrees with the decay rate
in Dirac's exponential decay and it has the following form:

\begin{equation}
A  = (e^2/3\pi\epsilon_{0}c^3\hbar) [(E_{2} - E_{1})^3/\hbar^3]
(\int \psi_{1} \textbf{r}\psi_{2} dr)^2
\end{equation}

where $E_{2}$ and $ E_{1}$ are the eigenergies of $\psi_{2}$ and
$\psi_{1}$  respectively.

After the successful calculation of the emission line width in
exponential decay by Weisskopf and Wiegner in 1930 on the basis
of Dirac's model of quantized light field and after Weltons
succesful interpretation of Lamb shift as a direct observable
effect of vacuum fluctuations in 1948, the folk theorem ``vacuum
fluctuations cause the spontaneous emission" established itself
in text books (see Baym (1969) p. 276), and radiation reaction
became more and more to be regarded rather as a classical concept
that is superfluous in quantum mechanics. In the fifties and
sixties, semiclassical models (par. \textbf{3.2.d}) in connection
with masers became popular again because the Dirac's transition
probabilities were not sufficient for describing phenomena where
phase relations between radiation field and atom's wave function
played an important role because of long term coherences. In these
semiclassical models, one tried  to find sufficiently accurate
solutions of the Schr\"{o}dinger equation for the given boundary
conditions.(Shimoda et.al. (1956)). Jaynes at al. enhanced
semiclassical theory by including the effect of source field back
on the electromagnetic field in the cavity and called it
neoclassical theory (NCT). What they wanted to do was not merely
to develop a better description of cavity phenomena, but to
challenge the established ``probability amplitude" interpretation
and to restore Schr\"{o}dinger's realistic interpretation of wave
function. The model succesfully described the cavity dynamics.

\subsection{Correspondence and discrepancy between nonlinear NCT
and linear QED}

The success of NCT led to the following question: ``What is the
degree of correspondence or discrepancy between the nonlinear
neoclassical equations and linear QED?" This was the question
that was addressed by Jaynes and Cummings in the (1963) paper.
They calculated the time evolution of an atom's dipole-moment and
electric field in a resonant cavity using the two competing
models, namely according to NCT based on Schr\"{o}dinger
interpretation of the wave function on one hand and pure quantum
mechanically using the commutators in Heisenberg picture on the
other hand.

They showed that the time evolution of the operators in
Heisenberg picture agreed to a great extent with the time
evolution predicted by coupled nonlinear semiclassical equations
(p.101 fig.4) even in the few photon region. (p.
97)\footnote{There is a typographical error on page 97 regarding
the equation number referred in the text. In the phrase
\textit{"...which is to be compared to (12c). If we interpret
(12c)as the expectation value of (20),...."} the authors refer in
truth to (69c) instead of (12c) and to (77) instead of (20)}. The
main difference between QED and NCT seemed only the lack of
correlated states in NCT. In NCT product of expectation values
appear where the expectation value should be taken after the
multiplication of the operators according to QED. The actual time
evolution was differing only slightly in the two models(p.101
fig.4). Thus calculations indicated that Schr\"{o}dinger
interpretation was not as in contrast to QED as previously
thought. Thus Fermi's nonlinear model of spontaneous emission (8)
is not a mutually exclusive alternative to Dirac's description,
inconsistent with principles of QM  but it is an incomplete
approximate model that reflects only a partial aspect of the
process.

In 1969 Crisp and Jaynes calculated lamb shift and spontaneous
emission using the semiclassical model where the lamb shift and
spontaneous emission appear merely as a consequence of radiation
reaction. In 1970 Stroud and Jaynes presented an improved
semiclassical model taking the results of 1969 paper into
account. There were only some small differences between
predictions of QED and the NCT.

\subsection{Source field in perturbative QED}

The success of NCT  led to controversies because of the
fundamental difference in the interpretation of wave function
when compared with QED. This success of NCT motivated QED
defenders to take a closer look at the mechanism of spontaneous
emission. Based on pure QED calculations, Ackerhalt et.al.(1973)
suggested that radiation reaction should play the dominating role
in spontaneous emission.  Detailed calculations later showed that
one could see vacuum fluctuations and radiation reaction as two
faces of the same reality and the interpretation one may adopt
depend on the ordering of the operators. (Milonni et.al (1973);
Senitzky (1973); Milonni Smith (1975); Milonni (1976)). Milonni
(1976) (1980) suggested  that spontaneous emission should be
considered as a result  of the combination of both effects.
Milonni showed also that although the interpretation of the cause
of line shift may depend on ordering of the operators, the role of
radiation in dynamics of energy loss does not depend on ordering.
This combined role of vacuum fluctuations and radiation reaction
in spontaneous emission is widely accepted today. (Dalibard et.
al. (1982)(1984); Cohen-Tannoudji (1986)).

Despite the great success of NCT, there were early signs of
conflicts with experiments regarding correlated outcomes in
measurements (Kocher (1967), Clauser(1972)). There were also
discrepancies regarding emission line shape. As time passed and
experiments improved, QED won the battle over NCT on these
issues\footnote{ except one prediction of NCT defenders regarding
quantum beats in $\Lambda$ type atoms that we will discuss in
paragraph 5.f in section 5. This was however not the deficiency
of QED itself but the misinterpretation of QED by its
defenders.}. However NCT served a great purpose in the history of
physics: By stimulating the reconsideration of source field
effects in QED, it helped the physics community to realize that
the consideration of source field effects does not necessarily
need semiclassical assumptions and it stimulated the efforts to
find solutions to source field problem beyond perturbative QED.

\subsection{Source field dynamics beyond perturbative approach}
In all  these discussed QED models, the potential energy of the
dipole is considered in the Hamiltonian  by a potential term of
the form $1 /(r_1 -r_2)$  and not as an integral over $E^2$. It
is only the free radiation field that is expanded in terms of
creation and annihilation operators. The role of radiation
reaction can be identified only by the interpretation of the
dynamics after appropriate ordering of operators. However we know
that a potential term in form $1 /(r_1 -r_2)$ has only approximate
validity because it implies that the Coulomb field immediately
changes over the whole space when the position of the particle
changes. This is not true from the relativistic point of view and
one has to take the retardation into account. That energy is
transported away by electromagnetic radiation indicates that the
energy is indeed in the electromagnetic field and not in the
charges so that we have to write it as an integral over $E^2$.
The following example shows that  we cannot escape this: Consider,
for example, a charged particle with mass $m_1$ in the
gravitational field of a neutral massive object with mass $m_2$.
The hamiltonian must contain a gravitational potential term of the
form $g m_1 m_2/(r_1 -r_2)$, but there is no electrostatic
potential term of the type $1 /(r_1 -r_2)$ because there is only
one charged particle. The particle has obviously a Coulomb field
however unsharp its value is according to QM, and it radiates
because of acceleration as it is assumed to happen when x-rays
are emitted while plasma from neighboring star is falling into a
black hole. There are no electrical dipoles in synchrotron
radiation either. Thus, we must somehow find a way to describe
radiation reaction quantum mechanically even for a single
particle wave packet independent of the nature of the force that
accelerates it.

Sharp and Moniz were aware of this problem and turned their
attention to a purely quantum mechanical nonperturbative
treatment of radiation reaction of a single particle ( Moniz and
Sharp (1974)(1977); Sharp (1980)).  Their starting point was the
following hamiltonian:

\begin{equation}
H = \frac{1}{2m}[\textbf{P} - \frac{e}{c} \textbf{A(r)}]^2 +
\frac{1}{8\pi} \int d \textbf{R} \{\textbf{E}^2(\textbf{R},t) +
[\nabla \textbf{ } \times \textbf{  A}(\textbf{R},t)]^2\}
\end{equation}
where
\begin{equation}
\textbf{A(r)} = \int d  \textbf{R } \rho [\textbf{ R -
r(t)}]\textbf{A}(\textbf{R},t)
\end{equation}
and
\begin{equation}
\textbf{E} = \textbf{E}_{long.} + \textbf{E}_{transv.}
\end{equation}

$\textbf{r}$ is the operator for particle coordinate and
$\textbf{R}$ is the general space coordinate. The electron is
assumed to have a spherical charge distribution $\rho$ so that $
\int \rho( \textbf{r - R})= 1 $ . Thus, unlike the potential term
in the hamiltonian of a two particle system, the Coulomb energy in
the field  is written as an integral over the field.  They
obtained the operator form of the Lorentz equations in Heisenberg
picture.  As a result of rigorous calculations, quantum mechanical
equivalent of the second term in the operator equation
corresponding to the classical equation (7) automatically appears
on the right side plus some purely quantum mechanical smaller
terms with higher powers of $c$ in denominator.  If one takes the
point charge limit at an appropriate stage, it turns out that
additional purely quantum mechanical smaller terms generate an
effective charge distribution spread out over a compton wave
length. This eliminates automatically the unphysical runaway
solutions (Moniz and Sharp (1974); Levine et al (1977). If we take
the expectation values of the operator $\textbf{r}$ on both sides
we obtain the desired modified form of eq.(5). Rohrlich
(1980)emphasizes the importance of their results in context with
fundamental problems of QED. Grotch et.al.(1982) discussed the
relation between perturbative approach and the work of Sharp and
Moniz  and emphasized the fact that perturbative approach can
account only for the effects related to retardation between two
particles and the contribution of self retardation cannot be
accounted for because of renormalization.

\subsection{Present situation}
After the work of Sharp and Moniz the problem of radiation
reaction has been  recognized as a problem by itself in QED, and
several works on the subject appeared since then classically as
well as quantum mechanically (Jim\'{e}nez (1987); Lozada (1989) ;
Ianconescu et.al.(1992) ; Ford et.al (1991) (1993); Kim (1999);
de Parga et.al (2001)). There is still no general consensus and
discussions continue (Rohrlich (2001)(2002) Ribari\u{c} (2002)
Baylis (2002) O'Connel (2003) Heras(2003)). Unfortunately
discussions on the quantum measurement problem seem to proceed
without taking all these developments into account despite the
relevance as mentioned in the introduction. Lets summarize the
possible reasons for this.

\section{Why has it not been considered in context with the
collapse problem until now?}

In the light of the above discussion, the possible reasons can be
summarized as follows:

\textbf{4.1.} The structure of QM and QED is  linear. Thus it was
assumed that any effect that can be described within the framework
of QED cannot be a candidate to explain the wave function collapse
and that one has to undertake modifications like making it
slightly nonlinear (Weinberg (1989)(1993)) or adding ad hoc
stochastic terms to achieve this.

In QED the Hamiltonian does not contain only the energy of the
particle, but also the energy of the electromagnetic field; and
the wave function in QED is not ``particle wave function" of the
ordinary QM anymore, but a more abstract wave function that
describes the whole system consisting of material particle(s) and
electromagnetic field. Although the linearity and superposition
principle still holds for the time evolution of this whole
abstract wave function, what we measure in a measurement is a
particular variable  like the position of the ``particle" or
frequency of emitted ``radiation" etc. Therefore, we have to focus
on the dynamics of the subsystem like, for example, ``wave
function of the measured particle". However, such an exact
separation of the total wave function is not possible in QED
(although we will use the term ``\textit{particle wave function}"
in  approximate sense). Nevertheless we can use Heisenberg
picture and focus on time evolution of the operator corresponding
to the particular variable of interest during the interaction of
radiation with matter. As the work of Jaynes Cummings have shown,
the back coupling of the atom's source field to radiation field in
the cavity leads to effective nonlinearity of time evolution of
the atomic operator.

Thus, if one focuses on time evolution of a subsystem like
``particle wave function" as we implicitly do in a discussion of
quantum measurement, then the linearity of the QED is a wrong
argument to principally discard the possibility that the dynamics
of the collapse can be described within laws of QED.

Whether the destruction of superposition of "particle wave
function" by radiation reaction proceeds sufficiently fast to
amplify initially small differences rapidly enough, to  explain
the collapse is still an open question that requires further
research to clarify, in particular it requires the consideration
of the results of theoretical papers about quantum mechanical
description of radiation reaction in measurement theories.

\textbf{ 4.2.} Radiation reaction was considered for long as a
semiclassical concept. It was believed that its consideration
necessarily uses implicitly the mental picture ``$|\Psi|^2=$
classical charge density"  that is inconsistent with fundamentals
of QED where $\Psi$ should be interpreted only as a probability
amplitude but not as something physically real that carries the
charge (see paragraph 5.f in the next section). Thus it was
assumed that the dynamics of this probability wave could not lead
to some real physical event. The role of radiation reaction  in
QED description was realized relatively late, namely in the
seventies in connection with spontaneous emission. Not only was
Copenhagen interpretation already well established at this time
but the skeleton of the most of the well known alternatives to
Copenhagen interpretation (like Bohmian mecanics, Everett's
relative state interpretation, von Neumann approach that led to
decoherence theories etc.) had already been formed and started an
evolution of their own.

Thus, the situation is that, on one hand scientists working on
decoherence approach to quantum measurement seem not sufficiently
familiar with all this complicated aspects of radiation theory
related to source field effects. On the other hand, scientists
working in radiation theory in general and on quantum mechanical
description of source field effects in particular have not
explored the consequences of their calculations for the
measurement problem. Thus what is needed seems to be an
inter(sub)disciplinary cooperation.

\textbf{ 4.3.} The calculation of source field effects starting
from first principles is not an easy task. Calculations are
already complicated enough in a nonrelativistic treatment as the
work of Sharp and Moniz show. Most of the papers on the subject
that followed this work use ad hoc introduction of a form factor
to avoid infinities. The appealing property of the work of Sharp
and Moniz is that it doesn't rely on ad hoc introduction of finite
particle size. Therefore, it can be considered as a milestone on
the subject and calculations  must be extended into the
relativistic region. Discussions on the quantum mechanical
description of radiation reaction continue and there is still no
general consensus on the subject. The lack of general consensus
and the complexity of the calculations even for a single particle
prevents its application to more complicated problems like
interaction with a measuring device. It seems the solution
requires a numerical approach and computer simulations.

\textbf{4.4.} Compared to other radiative phenomena, radiation
reaction is considered as a weak process that can be ignored.
Although this is true for a typical transition between two atomic
bound states, it is not generally true as one can see in formula
(9) that gives the magnitude of radiation reaction for a
transition between two stationary states. Even for slightest
energy differences($E_2 - E_1 $) that can emerge by a slightest
removal of degeneration between involved states by fluctuations,
large spatial extension of the involved states (as we have when
the particle wave function with large transversal extension enters
the measuring device) may lead in general to large transversal
dipole moment matrix elements $(\int \psi_{1} \textbf{r}\psi_{2}
dr)$ . This large dipole moment may lead to rapid decay rate
towards states that are more stable with regard to radiation
reaction(namely more stationary or more localized). Consider also
that it is not only the dipole moment that contributes to decay
but the third time derivative of all position operator matrix
elements that contribute to radiation reaction in the case of a
spatially extended wave function according to nonperturbative QED
calculations of Sharp and Moniz.

So far we have discussed the detection of material particles. We
want to make a short remark about the detection of photons. Based
on the considerations above one may suggest that simultaneous
excitation of multiple detector atoms by a single
photon\footnote{Here we use the term photon in the field
theoretical sense namely ``n photon state" referring to the ``n'th
energy eigenstate of the field" in a particular vibration mode
and not in the sense used in wave particle duality of old QM } is
not principally impossible because there is a
particle-like(pointlike) aspect of the photon complementary to its
wave like aspect\footnote{Consider the excitation of vibrational
modes of a water molecule in microwave range. The whole molecule
as a single multiparticle quantum system absorbs the energy of a
single photon over many cycles. We cannot say that the incoming
photon is absorbed by the oxygen atom or by the hydrogen atom.
There is no evidence that there is a pointlike entity that
disappears suddenly at some point during detection(absorption)}
but it is merely extremely unlikely because if  two distant atoms
are halfway excited by a single photon, both atoms are in a
nonstationary superposition so that such a state would be instable
because of source field effects.

\section{Discussion of general arguments against dynamical collapse
or against physically real wave function}

Based on whatever mechanism, if you can fully describe with your
deterministic wave equation why only one of the possible outcomes
is observed  you don't need to attach a special ambiguous
ontological status of ``probability amplitude" to the wave
function. The observed apparently probabilistic behaviour is then
only a consequence of dynamics but not an inherent property of
the wave function itself. We must note that this doesn't only
apply to a real collapse towards  a unique outcome as we suggest,
but it applies also to the decoherence picture where the wave
function is ``sliced" into noninterfering outcomes. In
decoherence picture, for example, the pointlike detection of an
electron on a scintillating screen at a particular position does
not indicate the existence of a pointlike aspect complementary to
wavelike aspect but it represents only one of the decohering
branches of the detector states corresponding to detection at
different locations.(see paragraph on decoherence in subsection
1.1). Thus, there is no need for a pointlike entity to understand
pointlike scintillations and thus no wave particle duality if we
accept the logical consequences of any type of dynamic collapse
within laws of QED.

Obviously a succesful deterministic dynamic collapse theory within
QED would explain how the magnitude of the wave function makes the
corresponding outcome more probable. This would make the
probability amplitude interpretation superfluous and impose
automatically a realistic interpretation of the wave function. If
we suggest such a possibility in principle we have to discuss in
advance general arguments (\textbf{5.a }to \textbf{5.g} ) that
have been used against realistic interpretation and against
dynamic collapse.

\textbf{5.a} ``\textbf{$\Psi$ is complex. Real physical entities
should be represented by real numbers}" The complex value
indicate that we have not only a magnitude but also a cyclic
property we call phase.  The absolute value of the phase has no
physical meaning. One can rotate the phase everywhere at the same
amount and the physics remains the same.  Consequently, the
absolute values of real and imaginary parts don't have physical
meaning. However, we don't see any philosophical reason why even
this should not be possible in nature. See arguments under
\textbf{5.b } regarding the relationship between observed reality
and deeper reality.

\textbf{5.b} ``\textbf{For n particles the wave function is
defined in 3n dimensional configurations space but real physical
space is 3 dimensional and therefore the wave function cannot be a
physical entity}". It is surprising that the defenders of
Copenhagen interpretation are on one hand so ``open-minded" to
accept totally new concepts like probability wave or mutually
exclusive aspects that reveal themselves depending on performed
experiment (complementarity) that are alien to classical
understanding of nature and the concept of objective reality, but
on the other hand they consider this as one of the arguments
against a realistic interpretation. The elements of the ``deeper
reality" (described by operators and the wave function) have
neither to be similar to our macroscopic reality nor should they
be directly observable, but it is enough if we can deduce directly
observable macroscopic experience described by eq.(5) from the
equations that describe this ``deeper reality", however strange
this ``deeper reality" may appear when viewed within a conceptual
framework built upon our daily macroscopic experience. This is why
a physical theory is not merely an empirical recipe to correlate
collected data but it represents a ``deeper insight" about how
nature works. The correspondence in form of eq.(5) does not only
work for a single particle but it works also for n particle
system.

\textbf{5.c} \textbf{``Quantum mechanical wave packets spread in
time but macroscopic objects do not."} If one considers, for
example, the two particle system of the hydrogen atom, the spatial
wave function is in 6 dimensional configurations space and one
can separate the Schr\"{o}dinger equation into two equations, one
with  the relative coordinates $r_r = r_p -r_e$ and the reduced
mass, and the other one with center of mass coordinates and the
total mass representing the center of mass motion of the whole
atom as a free particle. What we experience as size of an
hydrogen atom is the spatial extension of the ground state in
relative coordinate subspace, but this is of course independent of
the wave length or size of the wave packet corresponding to the
center of mass motion. It is the type of the involved interaction
that determines which part reveals itself in the experiment. What
we experience as the size of a macroscopic object in daily life is
the spatial extension of the bound state of the n particle system
in relative coordinate subspace. This is why macroscopic objects
do not spread. The free wave packet corresponding to center of
mass motion of a macroscopic (n particle) object is simultaneously
present in a different subspace of the 3n dimensional
configurations space.

\textbf{5.d} \textbf{ ``One can never observe wave function in its
entirety directly. What we can observe/detect directly are only
localized entities (particles). We can only calculate the wave
function by evaluating a large number of measurements conducted on
identically prepared particles. Thus, we don't have direct
observational evidence for the physical reality of the wave
function"} In contrast to the interference pattern on the
scintillating screen in a double slit experiment that emerges as
a result of large number of individual pointlike detections, the
interference pattern in the Bose Einstein condensate (Andrews
et.al (1997)) continue to exist however we lower the number of
the atoms in the condensate,(as long as we keep the temperature
deep enough) and we don't encounter pointlike entities
representing individual atoms in the condensate \footnote{do not
confuse this with grained structure in the photograph related to
individual detection of reflected photons. We are not talking
here about the wave function of detected photons but about the
condensate namely the ``entity" that reflects photons}. Indeed,
any localization of the individual atoms would mean the
destruction of BE condensate. In the double slit interference
experiment we can not observe the electron while passing as a
wave through both slits simultaneously because electron wave
collapses due to interaction with photon to a localized state.
Therefore, it passes either through one or the other slit so that
we loose the interference pattern. In the case of Bose Einstein
condensate light is reflected from condensate to the camera
without destroying its coherent quantum state as long as the
temperature remains sufficiently low. Thus the photographs of
interference patterns in a Bose Einstein condensate are in a
sense direct observation of the wave function of the condensate
without collapse. It seems that although the wave function is more
strange than (or less similar to) a classical field in ordinary
three dimensions as Schr\"{o}dinger originally hoped, it is
nevertheless physically more real then the defenders of the
Copenhagen interpretation had assumed.

\textbf{5.e} \textbf{``The form of the potential energy operator
for two particles has the form $1/(r_1-r_2)$ . If the wave
function were something physically real then we would have to
write the potential energy term as integral between two continuous
charge distributions. This is however not the case."} This
argument implicitly assumes that a realistic interpretation is
automatically equivalent to Schr\"{o}dinger interpretation.  The
error in this assumption is reading the correspondence principle
in a wrong direction by assuming that $|\Psi|^2$ must be something
like \textbf{classical} charge density. The correct reading of
correspondence principle is ``what we experience as classical
charge density in ordinary 3 dimensional space is the result of
the form of the wave function in 3n dimensional configuration
space that evolves according to \textbf{purely quantum laws}".

One should not confuse between two opposite directions of
correspondence between classical physics and quantum mechanics,
namely

\textit{ i.} How we discover the form of the operator in quantum
mechanics by looking at the corresponding classical expression

\textit{ ii.} How classical laws emerge as a consequence of
quantum laws as in equation (5)

It is obvious that only one of them can be fundamental so that
the other one must be the logical consequence of the more
fundamental one. Obviously, what counts is not how we discover the
laws but how nature works. The nature works according to \textit{
ii} not \textit{ i}. The reason why two localized quantum
mechanical wave packets move in first approximation similar to
classical bodies with a classical potential term $1/(r_1-r_2)$ is
because the quantum mechanical operator for potential term has
the form $1/(r_1-r_2)$ and not the other way round\footnote{It is
only our luck that it works in the other direction in such a
simple form as in \textit{i}}. Since it is only the dynamics of
the continuous entity wave function (or operators in Heisenberg
picture) what QM is about, the mathematical form of any quantum
operator can not be an argument for the existence of a pointlike
entity.

The correspondence in form of \textit{ii}  can be confirmed
experimentally in Rydberg atoms, where one can excite the
electrons wave function  to arbitrarily sharp localized wave
packets in an elliptical orbit by adjusting the shape of the
laser pulse accordingly. All these support the assumption that
localized detections may be  a spontaneous emission type rapid
(3.2) but continuous transition from a spatially extended state
to a localized one\footnote{localized as a consequence of the
detector potential not as an intrinsic aspect of the measured
entity} and not a "\textit{sudden revealing of particle aspect
that remains hidden until it is measured}".

\textbf{ 5.f.} \textbf{`"Neoclassical theory (NCT) was developed
on the basis of a realistic interpretation of the wave function.
It was not in agreement with experiments regarding emission line
shape and correlations between individual measurement results.
This shows that the realistic interpretation is inconsistent with
observations"}

The problem of semiclassical or neoclassical model was not the
realistic interpretation of particle wave function but
associating a classical charge density with it and rejecting the
operator nature (quantization) of the EM field.  Thus the error
was the wrong reading of the correspondence principle we mentioned
in first paragraph under \textbf{5.e} The resulting difference
between QED and NCT as a consequence of this is that in NCT
product of expectation values appear where the expectation value
should be taken after multiplication of operators according to
QED.  There is nothing in equations of QED that says wave
function is only probability amplitude and it cannot be something
real. It is only  the speculation ``collapse can principally not
be described as a dynamical process within QED" that leads to the
assumption that probability amplitude interpretation is an
inseparable part of QED. A good example of how this assumption
could mislead even the experts is the discussion on quantum beats
in $\Lambda$ type atoms\footnote{A three level atom that is
excited to the upmost level initially and decays spontaneously to
the lower levels. NCT defenders expected variation of the emitted
intensity due to interference between two concurrently occurring
transitions.} in the seventies that couldn't be observed at that
time(Jaynes (1980), Scully (1980). QED defenders claimed that
expecting beats in $\Lambda$ type atoms is based on the realistic
Schr\"{o}dinger interpretation and the missing of the beats
verify the invalidity of it and confirm probability wave
interpretation. The beats in $\Lambda$ type atoms were actually
observed in 1995 (Schubert et.al (1995)).

All problems mentioned above are, therefore, not real problems in
our opinion; however it seems the following  problem is a serious
one that has to be solved.

\textbf{5.g} \textbf{``Correlations between space-like separated
measurement outcomes conducted on entangled particles point to a
conflict with special relativity if we want to describe collapse
as a rapid but continuous dynamic process occurring at the very
moment of detection."} (Aspect et.al (1980)) Although there are
still discussions about possible loopholes, each and every
improved experiment seems to continue to confirm the violation of
Bell inequalities(Chiao et.al (1993), Tittel et.al (1998), Chiao
et.al (2002)). The usual reconciliation of nonlocal correlations
with special relativity goes as follows(summarized by us)(see for
example Chiao et.al (1993)): \textit{``Each individual
measurement on a single particle is random. This is independent
of whether the particle is entangled with another particle or
not. Thus the measurements on an individual particle do not
reveal any information about entanglement with another particle.
The correlations become apparent only after the measurement
results are brought together or communicated via usual means with
subluminal speed.  Therefore, there is no superluminal information
flow"}. Fundamental randomness is a necessary condition for such
a reconciliation. Thus denying the ontological status of the wave
function (Stenger (1995) p. 196) and interpreting it
\textit{only} as a probability amplitude, namely merely as a
mathematical function determining the probability of an outcome
is the essence of the reconciliation. Any deterministic mechanism
\footnote{in our opinion including decoherence contrary to the
claims of decoherentists because decoherence is a local process
and there is no mechanism that assures that space-like separated
correlated outcomes remain in the same branch of history.}would
invalidate it.

How can we resolve the conflict with special relativity? Is there
something going on similar to the propagation of electromagnetic
waves in matter with superluminal phase velocity? Or is the
process similar to faster than light motion of wave packet maxima
(Friedrich (1995)) despite subluminal group velocity? Is it
because the continuity equation does not hold for each subsystem
(particle or electromagnetic field) separately (because an exact
separation of wave function is not possible) but it holds only
for some type of total density with subluminal currents so that
we may have a change of\textit{ particles $|\Psi|^2 $} without a
corresponding current\footnote{nonrelativistically $\Psi^* \nabla
\Psi - \Psi \nabla \Psi^* $ in the absence of electromagnetic
field}? Does the relativistic form of the operators automatically
assure that quantummechanical density currents remain subluminal?
Is it possible that Lorentz invariance of the equations is lost
because of the necessary modifications we have to make to
describe the source field effects?  Some of these alternatives
may sound dangerous at first sight since they seem to jeopardize
the established relativistic view of space time but we must bear
in mind that Lorentz transformations (LT) have two different
interpretations. In his original derivations of LT, Lorentz
considered the time dilation (Larmor dilation) and length
contraction (Lorentz-Fitzgerald contraction) as physical effects
due to a motion relative to a preferred frame\footnote{that could
be well understood by Maxwell equations}. The time dilation leads
to an inevitable time difference between  distant clocks
(symbolized by t' and called by Lorentz as ``local time") in the
moving frame when clocks are separated after synchronization
because of different absolute velocities in opposite directions
during separation. The time difference is the same, one obtains
if one tries to synchronize them by light signals in moving frame
because of different velocities of light in different directions
in moving frame. Thus it is impossible to check one method
against the other, and therefore there is no way to verify this
asynchronity. The reason why all the effects of time dilation and
length contraction seem to be reversed when viewed from moving
frame is essentially the fact that the moving observer is fooled
by assuming his distant clocks all to be synchronous clocks. A
Lorentzian derivation of LT can be found in Kennedy and
Thorndike's (1932) paper\footnote{The Lorentzian derivation is
given in the last two pages. Interestingly in some relativity
text books and on some web pages teaching special relativity it is
mentioned that ``Although Lorentz Fitzgerald length contraction
hypothesis could explain Michelson Morley experiment it cannot
explain Kennedy and Thorndike experiment". This creates the wrong
impression that Kennedy Thorndike experiments impose Einsteinian
interpretation and invalidates the Lorentzian view. What Kennedy
Thorndike experiment has shown is merely that the absolute length
contraction is not enough and one has to add an absolute time
dilation to explain the experiment. Kennedy and Thorndike derive
LT starting with these two assumptions as Lorentz had done it
previously. It seems that there is a general lack of information
about the degree of observational equivalence between these two
interpretations and about the reasons why Einstein's
interpretation was preferred. One example of this is the wrong
expectation of observation of a torque in Trouton Noble
experiment}. For Einstein the symmetry between considered
reference frames was a too precious a property to sacrifice its
physical reality by interpreting it as merely observational or
apparent. He sacrificed rather the established view about
absolute time and absolute simultaneity to save the interpretation
of the symmetry as physically real. Thus, in his interpretation
simultaneity is relative and the symmetry between the reference
frames is real and reflects the very structure of space time,
namely its Minkowskian type geometry. The Einsteinian
interpretation is preferred and has been established because of 3
reasons:

\textit{i.} It doesn't refer to a concept of undetectable
preferred frame, thus better in accordance with the principle of
Ockham's razor.

\textit{ii.} Symmetry is an appealing property for a fundamental
law.

\textit{iii.} From Lorentzian view it seems mysterious why the
velocity of electromagnetic waves should enter in \textit{all} the
fundamental equations, namely even the equations that seem to have
(at least at first sight) nothing to do with electromagnetism,
like for example gravitation or velocity dependence of the mass of
neutral particles etc. Einstein provides a simple answer to this
question. Since everything occurs in space-time and since the
geometry of space-time is Minkowskian, the physical equations
have to be Lorentz invariant.

However, despite the appealing properties of Einsteinian view,
Lorentzian interpretation may be worth beeing reconsidered in
context with the problem of quantum nonlocality as a last resort
because although  a superluminal speed is associated necessarily
with backwards flowing of time and leads to problems of causality
in Einstein's interpretation of LT, this problem doesn't exist in
Lorentzian interpretation where time is absolute. We must be
openminded enough not to raise the Einsteinian interpretation to a
taboo and not to principally discard the theoretical possibility
that the Lorentz invariance of the fundamental equations may not
be related to the geometry of space time but the reason may well
be that all different types of quants are merely different types
of excitations of the same field (similar to different phonon
modes in a crystal) in a galilean space time and that the
intrinsic fundamental properties of this unified field (like
elastic constants in a solid) reflects itself in dispersion
relations $\omega(k)$ of different type of excitations
\footnote{that we experience as relativistic energy momentum
relation $E(p)$} in form of rest masses of different type of
excitations  (namely $\omega$ at $k = 0$)and in form of an
asymptotical upper limit for group velocity $d\omega/dk $ for $k
\rightarrow \infty$ for all type of excitations. Although
Einstein's decision to sacrifice the established view of space
and time and to replace it by a relativistic space-time was
favorable because of the strong arguments; if the new upcoming
tough decision should be between either sacrificing
mathematically pleasing relativistic view of space time or
loosing the boundary between objective reality and our knowledge
on it, between physics and metaphysics so that we may be in
danger of loosing even the notion of objective reality itself it
should be the notion of objective reality that should be saved
from being sacrificed if science should continue to be science
and not turn into ``mediaval necromancy" as Jaynes called it in
1980 (p.43).

To address all these questions we must reach a consensus on the
quantum mechanical description of radiation reaction(see recent
references on radiation reaction in subsection 3.7). Then we have
to look at the Lorenz invariance of the obtained equations to
address problem \textbf{5.g}. Then we have to put the acquired
knowledge into the theories of quantum measurement. Probably it
would be appropriate to apply all these first to a simple case of
an electron wave packet of large spatial extension approaching two
attractive potential centers A and B to see whether small initial
asymmetries prevent it from evolving to a superposition of the
states localized at A and B so that it rather evolves almost
directly to the energetically more suitable one for large A-B
distance\footnote{By applying Fermi's semiclassical idea (eq.(8)
that ignores vacuum photons and considers merely radiation
reaction) to a three level atom (G\"{u}rel, A. ; G\"{u}rel,
Z.(1998))that is initially in the excited state $|3\rangle $ ,
one can actually demonstrate how the differences between initial
small contributions of the two lower states $|2\rangle $
$|1\rangle $ determine whether there is a transition directly
from $|3\rangle $ to  $|1\rangle $ or whether there is a cascade
in form $|3\rangle \rightarrow |2\rangle \rightarrow |1\rangle $.
A transition to a superposition of $|2\rangle $ and $|1\rangle $
occurs only if the initial contributions of $|2\rangle $ and
$|1\rangle $ have the same order of magnitude. This may explain
why the quantum beats for $\Lambda$ type atoms couldn't be
observed for a very long time until 1995 although they were
predicted by defenders of NCT in 1970's.}.

\section{Summary and conclusion}

Detection of a material particle is always associated with
radiation reaction. The consideration of source field dynamics in
the calculation of time evolution of particle position operator in
nonpertubative QED indicates that stationary particle states
remain unaffected by radiation reaction so that they are favored
in presence of a potential gradient(subsection 3.6 and 3.7).
Therefore it is necessary to investigate its possible role in the
destruction of superpositions of ``particle wave function" during
a measurement. Unfortunately, there are several historical and
technical reasons why radiation reaction is not discussed in
context with quantum measurement until now(section 4).

A realistic modeling of a quantum measurement that contains all
the complicated aspects related to radiation reaction seems
difficult analytically. In particular, there are still different
approaches to the problem of quantum mechanical description of
radiation reaction and discussions continue(see references in
subsection 3.7). We think that decoherence is not a solution to
measurement problem(subsection 1.1). The most important next step
needed in our opinion is the collaboration between scientists
that work on measurement theory and scientists who try to find an
appropriate quantum mechanical description of radiation reaction
so that the results can be applied to a sufficiently realistic
model of a simple measurement probably numerically. Contrary to
the opinion expressed by Fuchs et.al (2000) we think that a
realistic interpretation of all the quantum mechanical
entities(wave function , operators , vacuum fluctuations etc.)
would be the true ``QM without interpretation" and not the
established interpretation. Before exploring alternatives like
modifying Schroedinger equation (SE) by nonlinear or ad hoc
stochastic terms, it is important to be aware that SE as it is,
neither includes the effect of vacuum fluctuations that can be a
possible candidate for providing the necessary stochasticity
naturally from within QED, nor does it include the effect of
radiation reaction on particle wave function. Thus vacuum field
and source field dynamics may already provide the necessary
modification to SE naturally. In consideration of the principle
of Ockham's razor it is obviously preferable to look for a
solution within QED before considering ad hoc modifications to
SE. These alternatives should be explored  only if the
possibilities within QED are fully exhausted and proven to be not
a solution to the problem. One wonders whether it may be one of
the cases where the equations \footnote{equations of QED
including the description of source field effects} themselves
turn out to be smarter or the elements of the theory\footnote{wave
function,opperators} turn out to be more real then they were
suggested initially by scientists who discovered
them\footnote{Einstein added cosmological constant to the general
relativistic equations to prevent the universe from expanding in
the model but then removed it when it was discovered that
universe expands. He called this "the biggest error in my life"
}. \footnote{Dirac discarded negative energy solutions of his
equation as unphysical. It was later found that they describe
antiparticles}

\newpage
\large
 \textbf{References}\newline
\small

\noindent  Ackerhalt, J.R. ; Knight, P.L. ; Eberly, J.,
\textit{``Radiation reaction and radiative frequency shifts"},
Phys.Rev.Let., \textbf{30}, 456-460 (1973)\newline

\noindent Adler,S.L, \textit{"Why decoherence has not solved the
measurement problem:a response to P.W.Anderson"}, St.Hist. Phil.
Phys. \textbf{34}, 135-142 (2003)\newline

\noindent Andrews, M.R. ; Townsend, C.G. ; Miesner,H.J. ; Durfee,
D.S. ; Kurn, D.M. ; Ketterle, W., \textit{``Observation of
interference pattern between two Bose condensates"} Science 275 ,
637-641(1997)\newline

\noindent  Aspect, A.; Dalibard, J.; Roger, G.,
\textit{``Experimental Tests of Bell's inequalities Using
Time-Varying Analyzers"}, Phys. Rev. Let. \textbf{49}, 1804-1807
(1982)\newline

\noindent  Baylis, W.E; Huschilt,J physics/0204061 (2002)\newline

\noindent  Baym, G.\textit{ ``Lectures on quantum mechanics" }The
Benjamin/Cummings publishing company, inc. Reading,MA USA
(1969)\newline

\noindent Bohm,D.; Hiley,B.J. "The Undivided Universe"
Routledge:London (1993)\newline

\noindent Bub,J. "Interpreting the Quantum World", Cambridge
University Press:Cambridge(1997)\newline

\noindent  Chiao, R.Y., Kwiat, P.G ; Steinberg A.M.,
\textit{``faster then light" }Scientific American \textbf{269}
,52-60, August(1993)
\newline

\noindent  Chiao, R.Y. ; Kwait, P.\textit{ ``Heisenberg's
introduction of the ``Collapse of the wave packet" into quantum
mechanics"} (2002)\newline
http://xxx.lanl.gov/PS\_cache/quant-ph/pdf/0201/0201036.pdf\newline

\noindent  Clauser, J.F. \textit{``Experimental Limitations to the
validity of semiclassical radiation theories"}, Phys.Rev. A
\textbf{6}, 49-54 (1972)\newline

\noindent  Cohen-Tannoudji, C. \textit{``Fluctuations in radiative
processes"}, Phys. Scripta, T \textbf{12}, 19 (1986)\newline

\noindent Crisp, M.D ; Jaynes, E.T , \textit{``Radiative effects
in semiclassical theory"}, Phys. Rev. \textbf{179}, 1253-1261
(1969)\newline

\noindent  Dalibard, J. ; Dupont-Roc, J. ; Cohen-Tannoudji, C.,
\textit{``Vacuum Fluctuations and Radiation reaction
-Identification of their respective contributions"}, J.de
Physique, \textbf{43}, 1617 (1982)\newline

\noindent Dalibard, J. ; Dupont-Roc, J. ; Cohen-Tannoudji, C.
\textit{``Dynamics of a small system coupled to a reservoir"},
J.de Physique, \textbf{45}, 637 (1984)\newline

\noindent  de Parga, G.A. ; Mares, R. ; Dominguez
S.,\textit{``Lagrange-Gordayev method and the equation of motion
for a charged particle"}, Il Nuovo Cimento B \textbf{116}, 85-97
(2001)\newline

\noindent Dirac, P.A.M., \textit{``The quantum theory of emission
and absorption of radiation"},  Proc. Roy. Soc. A \textbf{114},
243-265 (1927)\newline

\noindent Everett,H. "Relative State Formulation of Quantum
Mechanics", Rev. of Modern Physics \textbf{29},454 (1957)\newline

\noindent Friedrich, H. ,\textit{``Faster than light motion of
wave packet maxima in a nondispersive medium"}, Am.J.Phys.
\textbf{63}, 183-184 , (1995)\newline

\noindent  Ford, G.W. ; O'Connell, R.F., \textit{``Radiation
reaction in electrodynamics and the elimination of runaway
solutions"}, Phys.Let.A \textbf{157}, 217-220 (1991)\newline

\noindent  Ford, G.W. ; O'Connell, R.F., \textit{``Relativistic
form of radiation reaction "}, Phys.Let.A \textbf{174}, 182-184
(1993)\newline

\noindent Fuchs, C.A. ; Peres, A. \textit{``Quantum Theory Needs
No interpretation"}, Physics Today , March, 70-71 (2000)\newline

\noindent G.C.Girhardi,G.C; Rimini,A.; Weber,T. "Unified Dynamics
For Microscopic and Macroscopic Systems",  Phys.Rev. D
\textbf{34}, 470 (1986)\newline

\noindent Griffiths,R. "Consistent Histories and the
interpretation of Quantum Mechanics", Journal of Statistical
Physics \textbf{36},219 (1984)\newline

\noindent Grotch, H.; Kazes, E. ; Rohrlich, F. ; Sharp, D.H.
,\textit{``Internal retardation"}, Acta Physica Austriaca
\textbf{54}, 31-38(1982)\newline

\noindent  G\"{u}rel, A. ; G\"{u}rel, Z.\textit{``Instability and
quantum indeterminacy"} (1998) \newline
http://physics-qa.com/html/qmapp.htm\newline

\noindent   Heras, J.A. , \textit{``The radiation reaction force
on electron reexamined"}, Phys.Let.A. \textbf{314}, 272-277 (2003)
\newline

\noindent  Ianconescu, R. ; Horwitz, L.P. \textit{``Self force of
a classical charged particle"}, Phys.Rev.A \textbf{45}, 4346-4354
(1992)\newline

\noindent  Itzykson,C. ;Zuber JB., S.\textit{ ``Quantum Field
theory"},McGraw-Hill , Singapore , (1993)\newline

\noindent  Jaynes, E.T. ; Cummings, F.W, \textit{``Comparison of
quantum and semiclassical radiation theories with application to
the beam maser"} Proceedings of the IEEE, 89-109 (1963)\newline

\noindent Jaynes, E.T.,\textit{``Quantum Beats"} in
\textit{``Foundations of radiation theory and quantum
electrodynamics"}  Ed. A.O Barut, Plenum Press New York, 37-43,
(1980)\newline

\noindent  Jim\'{e}nez, J.L. ; Campos, I. \textit{``A critical
examination of the Abraham-Lorentz equation for a radiating
charged particle"}, Am J. Phys. \textbf{55}, 1017-1023
(1987)\newline

\noindent Kennedy, R.J ; Thorndike, E. M. \textit{``Experimental
Establishment of the Relativity of Time"}, Phys. Rev. \textbf{42}
400-418 (1932)\newline

\noindent  Kim, Kwang-Je, \textit{``The equation of motion of an
electron: a debate in classical and quantum physics"}, Nuclear
Instruments and Methods in Physics Research A \textbf{429}, 1
(1999)\newline

\noindent  Kocher, C.A. ; Commins,E.D. \textit{``Polarization
correlation of photons emitted in an atomic cascade"},
Phys.Rev.Let. 18, 575-577 (1967)\newline

\noindent  Levine, H. ; Moniz, E.J. ; Sharp, D.H.,
\textit{``Motion of extended charges in classical
electrodynamics"}, Am J. Phys. \textbf{45}, 75-78 (1977)\newline

\noindent Lozada, A. \textit{"General form of the equation of
motion for a charged particle"}, J.Math.Phys. \textbf{30},
1713-1726 (1989)\newline

\noindent  Milonni, P.W. ;Ackerhalt, J. R.; Smith, Wallace A.,
\textit{``Interpretation of radiative corrections in spontaneous
emission"}, Phys.Rev.Let., \textbf{31}, 958-960 (1973)\newline

\noindent  Milonni, P.W. ;  Smith,W. A., \textit{``Radiation
reaction and vacuum fluctuations in spontaneous emission"},
Phys.Rev. A, \textbf{11}, 814-824 (1975)\newline

\noindent Milonni, P.W.  \textit{``Semiclassical and
quantum-electrodynamical approaches in nonrelativistic radiation
theory"}, Phys. Rep. \textbf{25}, 1 (1976)\newline

\noindent  Milonni, P.W , \textit{``Classical and quantum theories
of radiation "} in \textit{``Foundations of radiation theory and
quantum electrodynamics"} Ed. A.O Barut, Plenum Press New York ,
1-21 , (1980)\newline

\noindent  Moniz, E.J. ;  Sharp, D.H. , \textit{``Absence of
runaways and divergent self-mass in nonrelativistic quantum
electrodynamics"}, Phys.Rev. D \textbf{10}, 1133-1136
(1974)\newline

\noindent   Moniz, E.J. ;  Sharp, D.H. , \textit{``Radiation
reaction in nonrelativistic quantum electrodynamics"}, Phys.Rev. D
\textbf{15}, 2850-2865 (1977)\newline

\noindent  O'Connell, R.F., \textit{``The equation of motion of an
electron"}, Phys.Let.A\textbf{313}, 491-497 (2003)\newline

\noindent Ribari\u{c}, M. ;  \u{S}u\u{s}ter\u{s}ic, L. ,
\textit{``Qualitative properties of an equation of motion of a
classical point charge"}, Phys.Let.A  \textbf{295}, 318-319
(2002)\newline

\noindent  Rohrlich, F. \textit{``Fundamental Physical Problems of
Quantum Mechanics"} in \textit{``Foundations of radiation theory
and quantum electrodynamics"} Ed. A.O Barut, Plenum Press New
York, 155-163,(1980)\newline

\noindent  Rohrlich, F. \textit{``The validity limits of physical
theories:response to the preceeding Letter"},Phys.Let.A,
\textbf{295}, 320-322 ,(2002)\newline

\noindent  Rohrlich, F. \textit{``Dynamics of a clasical point
charge"},Phys.Let.A, \textbf{303}, 307-310 ,(2002)\newline

\noindent  Rohrlich, F. \textit{``The correct equation of motion
of a classical point charge"},Phys.Let.A, \textbf{283}, 276-278
,(2001)\newline

\noindent M.Schubert, I. Siemers, R. Blatt, W. Neuhauser,
P.E.Toschek "Transient internal dynamics of a multilevel ion"
Physical Rev. A vol 52 p. 2994 (1995).\newline

\noindent Scully, M.O.  \textit{"On Quantum beats phenomena and
the internal consistency of semiclassical theories"} in
"Foundations of radiation theory and quantum electrodynamics" Ed.
A.O Barut, Plenum Press New York (1980),pp.45 \newline

\noindent Sharp, D.H., \textit{``Radiation reaction in
nonrelativistic quantum theory"} in \textit{``Foundations of
radiation theory and quantum electrodynamics"} Ed. A.O Barut,
Plenum Press New York, 127-141,(1980) \newline

\noindent  Shimoda, K. ;  Wang, T.C.; Townes, C.H.
,\textit{``Further aspects of maser theory"}, Phys. Rev.
\textbf{102}, 1308-1321, (1956)\newline

\noindent Senitzky, I.R. \textit{``Radiation Reaction and
Vacuum-field Effects in Heisenberg-Picture Quantum
Electrodynamics"} Phys. Rev. Let. \textbf{31}, 955-958,
(1973)\newline

\noindent  Stenger, V.J.,\textit{ ``The unconscious quantum"}
Prometheus books New York (1995)\newline

\noindent  Stroud Jr., C.R.; Jaynes, E.T., \textit{``Long-term
solutions in semiclassical radiation theory"}, Phys. Rev. A
\textbf{1}, 106-121 (1970)\newline

\noindent  Tittel, W.; Brendel, J.; Gisin, B.; Herzog, T;
Zbinden, H.; Gisin, N. \textit{``Experimental Demonstration of
Quantum Correlations over more then 10 km"}, Phys. Rev. A
\textbf{57}, 3229-3232, (1998)\newline

\noindent  Weinberg, S.\textit{ ``Testing Quantum
Mechanics"},Annals of Phys. \textbf{194}, 336-386, (1998)\newline

\noindent  Weinberg, S.\textit{ ``Dreams of a final
theory"},Pantheon books NY , (1993)\newline

\noindent Weisskopf,V.F.;Wigner E.P., "Berechnung der natürlichen
Linienbreite auf Grund der Dirac'schen Lichttheorie" Z.Physik
\textbf{63},54(1930) and \textbf{65}, 18(1930)\newline

\noindent  Welton, T.A., \textit{``Some observable effects of the
quantum-mechanical fluctuations of the electromagnetic field"},
Phys.Rev., \textbf{74}, 1157-1167 (1948)\newline

\noindent  Wodkiewicz, K., \textit{``Resonance fluorescence and
spontaneous emission tests of QED"} in "\textit{Foundations of
radiation theory and quantum electrodynamics"} Ed. A.O Barut,
Plenum Press New York, 109-117,(1980)\newline

\noindent Zurek,W.H., \textit{``Environment Induced Superselection
Rules"}, Phys.Rev. D  \textbf{26} ,\linebreak 1862-1880
(1982)\newline

\noindent Dudley,J.M , A.M Kwan ``Richars Feynmans popular
lectures on quantum electrodynamics the 1979 Robb lectures at
Auckland university" Am J.Phys. g4 (6) June 1996 p.694

\end{document}